\documentclass[aps,prb,twocolumn,superscriptaddress,showpacs]{revtex4}
\usepackage{graphicx,bm,color,float,textcomp}
\newcommand\seo{$\rm SrEr_2O_4$}
\newcommand\sho{$\rm SrHo_2O_4$}
\newcommand\sdo{$\rm SrDy_2O_4$}
\newcommand\slo{Sr$Ln_2$O$_4$}
\newcommand\bimno{$\rm Bi_3Mn_4O_{12}(NO_3)$}
\newcommand\ErI{Er$_1$}
\newcommand\ErII{Er$_2$}

\newcommand\PRB[3]{Phys. Rev. B {\bf {#1}}, {#2} ({#3})}

\newcommand\PRL[3]{Phys. Rev. Lett. {\bf {#1}}, {#2} ({#3})}
\newcommand\JPCM[3]{J. Phys.: Condens. Matter {\bf {#1}}, {#2} ({#3})}
\newcommand\JPSJ[3]{J. Phys. Soc. Jpn {\bf {#1}}, {#2} ({#3})}

\newcommand\JACS[3]{J. Am. Chem. Soc. {\bf {#1}}, {#2} ({#3})}

\newcommand\afm{antiferromagnetic}

\begin{document}
\title{ Coexistence of the long-range and short-range magnetic order components in SrEr$_2$O$_4$}
\date{\today}
\author{T.J. Hayes}
\affiliation{Department of Physics, University of Warwick, Coventry CV4 7AL, United Kingdom}
\author{G. Balakrishnan}
\affiliation{Department of Physics, University of Warwick, Coventry CV4 7AL, United Kingdom}
\author{P.P. Deen}
\affiliation{European Spallation Source ESS AB, P.O. Box 176, SE-22100 Lund, Sweden}
\affiliation{Institut Laue-Langevin, BP 156, 38042 Grenoble Cedex 9, France}
\author{P. Manuel}
\affiliation{ISIS Facility, Rutherford Appleton Laboratory, Chilton, Didcot OX11 0QX, United Kingdom}
\author{L.C. Chapon}
\affiliation{ISIS Facility, Rutherford Appleton Laboratory, Chilton, Didcot OX11 0QX, United Kingdom}
\author{O.A. Petrenko}
\affiliation{Department of Physics, University of Warwick, Coventry CV4 7AL,  United Kingdom}
\begin{abstract}
Single crystal neutron diffraction reveals two distinct components to the magnetic ordering in geometrically frustrated \seo.
One component is a long-range ordered ${\bf k}=0$ structure which appears below $T_N = 0.75$~K.
Another component is a short-range incommensurate structure which manifests itself by the presence of a strong diffuse scattering signal.
On cooling from higher temperatures down to 0.06~K, the partially ordered component does not undergo a pronounced phase transition.
The magnetic moments in the long-range commensurate and short-range incommensurate structures are predominantly pointing along the [001] and [100] axes respectively. 
The unusual coexistence of two magnetic structures is probed using both unpolarised and XYZ-polarised neutron scattering techniques.
The observed diffuse scattering pattern can be satisfactorily reproduced with a classical Monte Carlo simulation by using a simple model based on a ladder of triangles.
\end{abstract}
\pacs{75.25.-j, 75.47.Lx, 75.50.Ee, 75.40.Mg}
\maketitle
\section{Introduction}
The geometric frustration of magnetic interactions frequently results in a highly degenerate ground state of the system.
In many frustrated systems such as those based on the the pyrochlore structure, kagome and garnet lattices, this degeneracy is often responsible for cooperative paramagnetism, in which the magnetic correlations remain short ranged and the system fluctuates continuously in the degenerate ground state manifold.\cite{Canals_1998_2000,Moessner_1998,Gardner_Rev_2010}
The degeneracy may be lifted, and a particular type of long-range order stabilised at lower temperatures, by thermal or quantum fluctuations (through the order-by-disorder mechanism). 
Long-range ordering in such systems is frequently non-trivial, may be incommensurate with the crystal structure and may not prohibit a disordered component remaining down to very low temperatures.
The coexistence of magnetic ordering and cooperative paramagnetism has been recently observed in, for example, the disordered system with the fluorite structure Tb$_3$NbO$_7$.\cite{Ueland_2010}
It is, however, unusual to observe two different types of magnetic order within a single system which does not show any signs of chemical instability or phase separation.
We report such an observation in the rare-earth oxide \seo, in which a long-range ordered magnetic structure with propagation vector ${\bf k}=0$ coexists with a shorter-range incommensurate ordering that has a pronounced low-dimensional character.

\begin{figure}[tb]
\includegraphics[width=0.9\columnwidth]{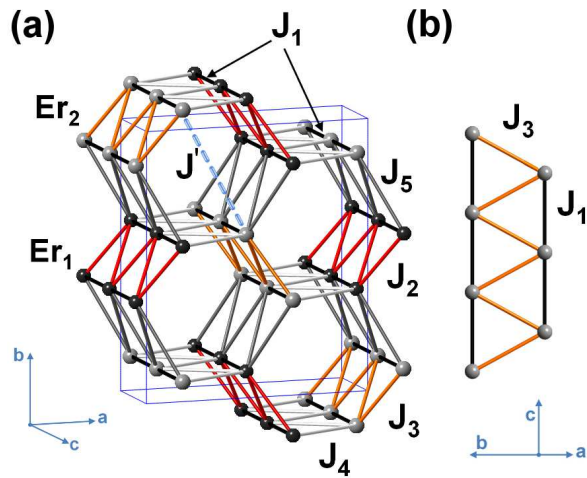}
\caption{(Colour online) (a) Positions of magnetic Er$^{3+}$ ions within the \seo\ crystal structure.
				Two inequivalent Er sites are shown as dark, \ErI, and light, \ErII, spheres.
				The chains of identical Er$^{3+}$ ions run along the \emph{c} direction, with the shortest distances, \ErI-\ErI\ and \ErII-\ErII, along the chains.
				They are coupled to parallel chains made from either the same or different Er$^{3+}$ sites.
				Different colours used for labelling the inter-chain bonds highlight the four distinct links between the Er$^{3+}$ ions.
				(b) Ladder of \ErII\ ions viewed along the $[210]$ direction (normal to the ladder) showing the nearest-neighbour interaction $J_1$ and the next-nearest neighbour interaction $J_3$.	
				}
\label{Fig1_structure}
\end{figure}

\seo\ belongs to the family of materials with the general formula \slo, where $Ln$ is Gd, Dy, Ho, Er, Tm, and Yb.
These oxides crystallise in the calcium ferrite structure\cite{Decker_1957} (space group $Pnam$),  in which the magnetic $Ln$ ions are linked through a network of triangles and hexagons\cite{Karunadasa_2005} (see Fig.~\ref{Fig1_structure}).
When viewed along the [001] direction, a honeycomb motif of $Ln$ ions is clearly visible, while an edge-sharing triangular motif is apparent when viewed along a direction perpendicular to the honeycomb planes.
Although triangular antiferromagnets are always subject to magnetic frustration, in a honeycomb lattice, where the magnetic ions form a 2D network of edge-sharing hexagons, nearest-neighbour interactions alone do not result in a frustrated system without competition from next-nearest-neighbour interactions.
Recent theoretical studies\cite{Mulder_2010,Okumura_2010} motivated by measurements on the \afm\ honeycomb system\cite{Smirnova_2009}~\bimno\ have not reached any firm conclusions as to the nature of the ground state in such a system with both nearest and next-nearest-neighbour interactions.

In the absence of reliable data on the strength or the sign of the interactions in \slo, it is not a priori clear whether the correct description of such systems is a 2D honeycomb, as quasi-1D chains, or a ladder consisting of triangles.
A system of similar geometry, $\beta$-CaCr$_2$O$_4$, has recently been shown to behave as a network of triangular zig-zag ladders.\cite{Damay_2010}
As will be shown below, our Monte Carlo simulations suggest that a simple model based on a ladder of triangles with a particular ratio of nearest and further neighbour exchange interactions captures adequately the main features of the magnetic diffuse scattering observed in \seo\ at low temperatures.

One important structural feature common to all \slo\ compounds is the presence of two inequivalent sites for the magnetic rare-earth ions in a unit cell (denoted as \ErI\ and \ErII\ for \seo\ as shown in Fig.~\ref{Fig1_structure}).
The identical ions are coupled to each other through the shortest distance, 3.39~\AA\, forming chains along the \emph{c} axis, the intra-chain exchange constant is labelled as $J_1$.
The chains of identical ions are then coupled into ladders by the exchange integrals $J_2$ (\ErI-\ErI\ separation is 3.48~\AA)  or $J_3$ (\ErII-\ErII\ separation is 3.51~\AA) shown in red and orange in Fig.~\ref{Fig1_structure}.
The ladders are separated from each other by the significantly longer distances of 3.87~\AA\ and 4.02~\AA.
The corresponding bonds are shown in Fig.~\ref{Fig1_structure} in grey and the exchange interactions are labelled as $J_4$ and $J_5$.
The lattice constants of \seo\ at 15~K are $a=10.018$~\AA, $b=11.852$~\AA, $c=3.384$~\AA\ (Ref.~\onlinecite{Petrenko_2008}).

Using powder neutron diffraction,\cite{Petrenko_2008} we have previously shown that \seo\ orders \afm ally at $T_N=0.75$~K.
This transition to an ordered phase is also marked by a $\lambda$-anomaly in the specific heat at $T_N$.
The ${\bf k}=0$ magnetic structure below this temperature was found to consist of ferromagnetic chains running along the \emph{c} axis, with adjacent chains arranged \afm ally.
The refinement suggested that the moments point along the \emph{c} direction, however, only one of the two Er$^{3+}$ sites possesses a sizeable magnetic moment (of 4.5~$\rm \mu_B$ at 0.55~K) contributing to this ordering.
Within the model constructed from this refinement, the magnetic moments may be swapped between the two Er$^{3+}$ sites without changing the calculated powder diffraction pattern significantly.

We have used high-quality single crystal samples for the current neutron diffraction study, allowing us to establish the presence of another magnetic phase which coexists with the previously reported structure over a wide temperature range.

\section{Experimental details}
Single crystals of \seo\ were grown by the floating zone technique using an infra-red image furnace, as previously described.\cite{Balakrishnan_2009}
The three different crystals used ranged in mass from 0.5 to 9.5 grams.
We have performed neutron diffraction experiments on single crystals of \seo\ using the PRISMA diffractometer at the ISIS pulsed neutron source in the UK and the D7 spectrometer at the ILL in Grenoble, France.
The PRISMA instrument~\cite{Harris_1998} utilised 16 $^3$He gas  detectors with a 1\textdegree\ separation providing a good $Q$ resolution, favourable signal to noise ratio and high neutron flux in the $Q$-range from 0.5 to 5.5~\AA $^{-1}$. 
D7 is a diffuse scattering spectrometer equipped with XYZ polarisation analysis,\cite{Stewart_2009} which uses 132 $^3$He detectors to cover the angular range $4^\circ < 2\theta < 145^\circ$ which when utilising cold neutrons monochromated to a wavelength of 3.1~\AA\ results in a $Q$-space coverage of 0.14 to 3.91~\AA$^{-1}$.
These experiments were carried out at temperatures between 0.05~K and 1.1~K utilising a dilution refrigerator inside a standard $^4$He cryostat.
The standard data reduction functions were used to create reciprocal space maps from the initial data and to normalise the detector efficiencies from vanadium standards. 
For polarisation analysis, a quartz standard was used to normalise the polarisation efficiency on D7.
Background scans were performed on D7 using equivalent sample mountings.

\section{Neutron Diffraction Results}
\begin{figure*}[tb]
\includegraphics[width=1.35\columnwidth]{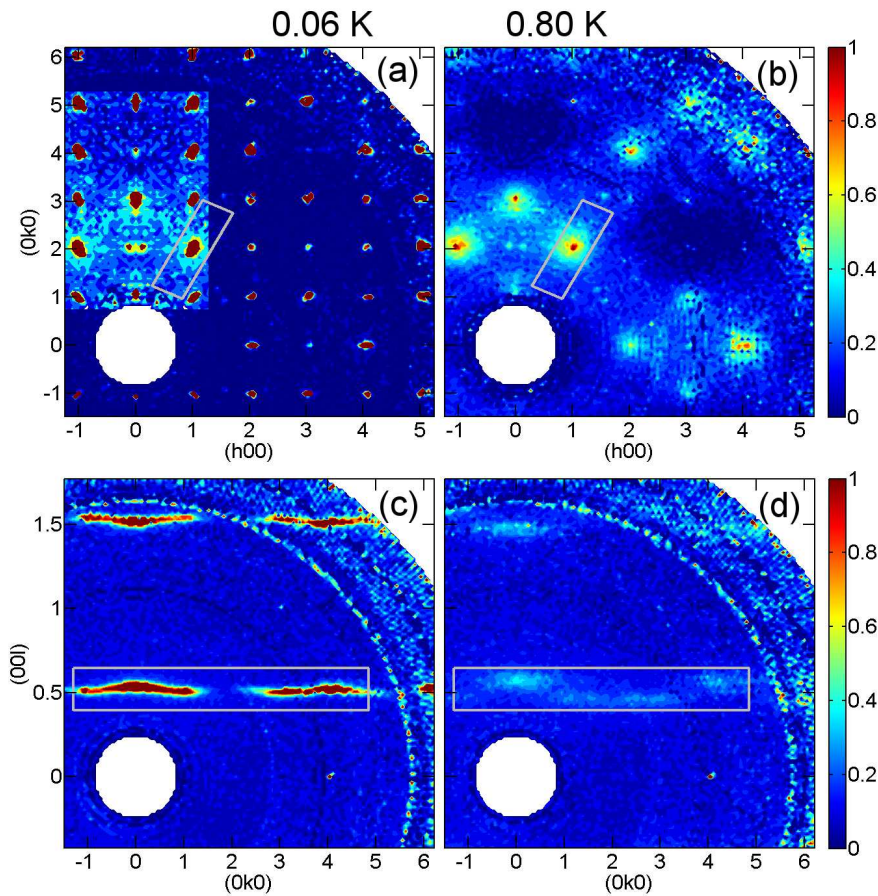}
\caption{(Colour online) Reciprocal space intensity maps of the magnetic scattering from \seo\ in the $(hk0)$ plane (top panels) and in the $(0kl)$ plane (bottom panels) at  0.06~K (left panels) and 0.8~K (right panels).
	      The highlighted area $-1.3<h<1.3$, $0.7<k<5.3$ in panel (a) has 10 times lower intensity scale to emphasise the presence at the lowest temperature of a weak diffuse scattering otherwise obscured by the much more intense Bragg peaks.
	      The magnetic scattering is isolated from the nuclear and spin-incoherent contribution by full XYZ polarisation analysis using D7 diffractometer for the $(hk0)$ plane.
	      In the case of maps of the $(0kl)$ plane, the intensity shown is obtained by removing the nuclear contribution from the non-spin-flip measurement with neutrons polarised orthogonal to the scattering plane,\cite{Zsub_expression} following Ref.~\onlinecite{Stewart_2009}.
	      The bounded boxes in all four panels delineate the areas of intensity displayed as cuts from the equivalent PRISMA data in Fig.~\ref{Fig3}.
	      }
\label{Fig2_maps}
\end{figure*}
The intensity maps shown in Fig.~\ref{Fig2_maps} summarise the main results of the neutron diffraction experiments.
In Figs.~\ref{Fig2_maps}a and~\ref{Fig2_maps}b the intensity in the $(hk0)$ scattering plane is shown for two temperatures: 0.06~K, the base temperature of the dilution refrigerator; and 0.8~K, just above $T_N$. 
At the higher temperature, broad diffuse scattering peaks are clearly present in positions occupied by sharp magnetic Bragg peaks below $T_N$.
Although these broad peaks are almost invisible in Fig.~\ref{Fig2_maps}a when using the same intensity scale, they remain present down to the base temperature.
A highlighted area limited by the conditions $-1.3<h<1.3$ and $0.7<k<5.3$ in Fig.~\ref{Fig2_maps}a has intensity ranging from 0 to 0.1 rather than from 0 to 1.0 as in the main panels. 
This area emphasises the presence of diffuse scattering (albeit with a significantly reduced intensity compared to the Bragg peaks) even at the lowest temperature.
This weak scattering forms a peculiar {\it lozenge} pattern, which is much more clearly seen at higher temperatures (as in Fig.~\ref{Fig2_maps}b) and which is mostly determined by the positions of the magnetic Er$^{3+}$ ions in the unit cell.

\begin{figure}[tb]
\includegraphics[width={0.9\columnwidth}]{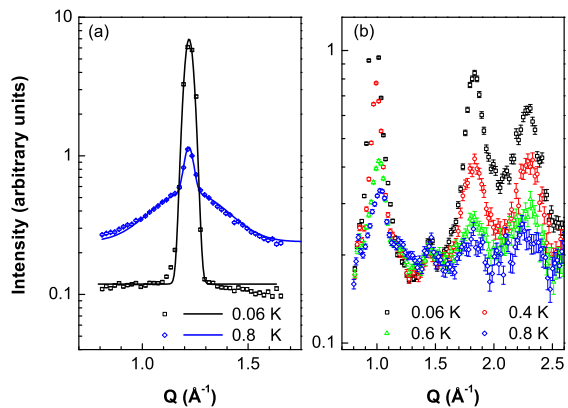}
\caption{(Colour online) Cuts through reciprocal space, with intensity plotted against $Q$, taken from experiments on the PRISMA diffractometer.
                                           (a) The (120) peak measured in the $(hk0)$ plane at the base temperature of 0.06~K and at 0.8~K (just above $T_N$).
                                           The fitted curves shown as solid lines are a single sharp Gaussian peak reflecting a strong magnetic scattering below $T_N$ and a combination of a sharp and broad Gaussian peaks reflecting nuclear contribution and magnetic diffuse scattering above $T_N$. 
                                           (b) The intensity measured along the $(0,k,\frac{1}{2}+\delta)$ feature in the $(0kl)$ plane at four temperatures.}
\label{Fig3}
\end{figure}

All of the sharp Bragg peaks observed in the $(hk0)$ scattering plane during the single crystal experiments are in agreement with the previously reported ${\bf k} = 0$ structure obtained from the powder refinement.
The corresponding magnetic structure may be found in Fig.~5 of Ref.~\onlinecite{Petrenko_2008}.
Example cuts through the $(hk0)$ plane at the (120) position are shown in Fig.~\ref{Fig3}a for the base temperature and 0.8~K, in which the diffuse scattering seen above $T_N$ is evident.
From the width of the fitted curve corresponding to the diffuse scattering, the correlation length is estimated\cite{FWHMnote} as $20 \pm 1$~\AA.

Similar measurements taken for the $(0kl)$ scattering plane reveal completely different diffraction patterns.
At the higher temperature of 0.8~K, broad diffuse scattering intensity forms clearly visible rods, which run along the horizontal $k$ axis and which are positioned around half-integer points along the vertical $l$ axis as shown in Fig.~\ref{Fig2_maps}d.
On cooling the system down to 0.06~K, these broad diffuse scattering rods became more intense (see Fig.~\ref{Fig2_maps}c).
As these rods extend along the $k$ direction, the magnetic moments are only weakly correlated along the $b$ axis.
The short-range magnetic correlations are either two-dimensional (confined mostly to the $ac$ plane) or one-dimensional (along the $c$ axis) in nature depending on whether the diffuse scattering is limited mostly to the $(0kl)$ plane or extends significantly out of this plane.
We will return to the discussion of this issue after presenting the results of Monte Carlo simulations in the next section.

As can be seen in Fig.~\ref{Fig2_maps}c, there is an intensity minimum around $k=2$, however, there is no pronounced decay of intensity with increasing $k$ or $l$ significantly more severe than that due to the magnetic form factor.
The lack of marked modulation in intensity of the rods with $l$ and $k$ is indicative of magnetic moments directed predominantly along the $a$ axis.
The positions of the intensity maxima along the $l$ axis are clearly incommensurate, however, their proximity to the half-integer points suggests that the unit cell for this magnetic structure is nearly doubled along the $c$ axis.
The shape of the rods also changes on cooling.
The two rods are at positions $(0,k,\frac{1}{2}+\delta)$ and $(0,k,\frac{3}{2}-\delta)$ respectively, where $\delta$ is dependent upon $k$, with the sign of $\delta$ alternating for each increasing integer $k$ value, and the magnitude of $\delta$ decreases upon cooling.
At the base temperature, the correlation lengths\cite{FWHMnote} along the $[001]$ direction for the rods, taken from the central $k=0$ position, are 128(2)~\AA\ for the $(0,k,\frac{1}{2}+\delta)$ rod and 70(4)~\AA\ for the $(0,k,\frac{3}{2}-\delta)$ rod.
Only at the lowest temperatures, when $\delta$ is smallest, do the rods begin to tend towards the perfectly straight line shape.
Fig.~\ref{Fig3}b shows cuts across the $(0kl)$ plane corresponding to $l = \frac{1}{2}$, containing one of the diffuse scattering rods, at 0.06~K, 0.4~K, 0.6~K and 0.8~K.
The positions and areas of these cuts are indicated in Fig.~\ref{Fig2_maps} as bounded boxes.
\begin{figure}[tb]
\includegraphics[width={0.7\columnwidth}]{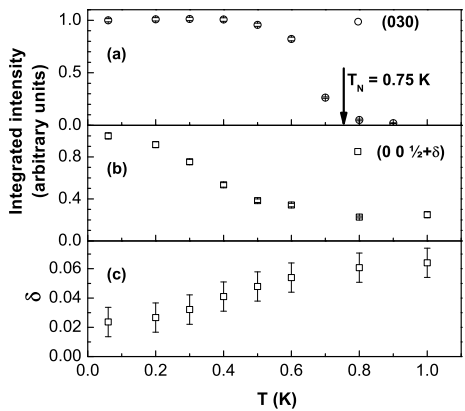}
\caption{Temperature dependence of (a) integrated intensity of the magnetic (030) Bragg peak as measured on the PRISMA diffractometer;
		(b) integrated intensity of the $(0, 0, \frac{1}{2} + \delta)$ broad feature limited to the region $|k|<\frac{1}{2}$ as measured in the $(0kl)$ plane on the D7 spectrometer;
		(c) the value of $\delta$.}
\label{Fig4_temperature}
\end{figure}

Indirect evidence for the existence of low-dimensional diffuse scattering at positions of wave vector $Q$ which cannot be accounted for in a ${\bf k}=0$ model is also found on re-examination of the powder diffraction results.\cite{Petrenko_2008}
In this data additional diffuse intensity is observed at 2$\Theta \approx$~21\textdegree and 2$\Theta \approx$~34\textdegree, corresponding to the most intense features seen in Fig.~\ref{Fig2_maps}c at the ${\bf k}=0$ positions of the two rods.  

The temperature dependence of the intensity of the (030) peak, associated with the ${\bf k}=0$ structure, clearly demonstrates the appearance of magnetic ordering below $T_N=0.75$~K, as shown in Fig.~\ref{Fig4_temperature}a.
Figs.~\ref{Fig4_temperature}b and \ref{Fig4_temperature}c on the other hand show that for the diffuse scattering observed in the $(0kl)$ plane no well-defined phase transition temperature is apparent.
For example, the integrated intensity of diffuse scattering measured in the $(0kl)$ plane for the $k < \frac{1}{2} $ region at $l \approx \frac{1}{2}$ gradually decreases with increasing temperature (see Fig.~\ref{Fig4_temperature}b), while the positions and shapes of the rods in reciprocal space, as characterised by the parameter $\delta$, are also smooth functions of temperature (see Fig.~\ref{Fig4_temperature}c).

\section{Monte Carlo Simulations}
In order to gain further insight into the intriguing physics of \seo, we have simulated the partially ordered magnetic arrangements and associated neutron scattering patterns of this compound via the Monte Carlo (MC) method\cite{Chapon_MC} using a standard Metropolis algorithm.
In addition to the isotropic exchange interactions, a single ion anisotropy term\cite{anisotropy} was introduced into the simulation with the \emph{b} axis being a hard direction of magnetisation as suggested by the magnetisation measurements.\cite{Petrenko_2008} 
The simulations were performed with either $10 \times 10 \times 30$ or $20 \times 20 \times 60$ supercells containing 12000 or 96000 Heisenberg spins with open boundary conditions.\cite{IQ}
Simulated annealing was performed from a temperature $T/J_1=400$ (to ensure the initial acceptance rate always stayed above 90\%) down to $T/J_1=0.001$ with an exponential cooling rate of 0.96 and 8000 spin flips per temperature.

Powder diffraction results\cite{Petrenko_2008} showed that only one of the two Er$^{3+}$ sites undergoes long-range magnetic ordering.
One can therefore conjecture that the observed low-temperature diffuse scattering is mostly due to the other site.
In the following simulations we presumed that the \ErI\ sites order while the \ErII\ sites do not.
An alternative arrangement where the \ErII\ sites order while the \ErI\ sites do not was also examined and gave matching results.
In the minimalistic model used to simulate the diffuse scattering, the magnetically ordered \ErI\ sites have been disregarded and only the \ErII\ ions and interactions between them are considered.
The \ErII\ sites form triangular ladders running along the $c$ axis that are parameterised with two interactions $J_1$ and $J_3$ along the legs and the rungs respectively as shown in Fig.~\ref{Fig1_structure}.
\begin{figure}[tb]
\includegraphics[width={0.7\columnwidth}]{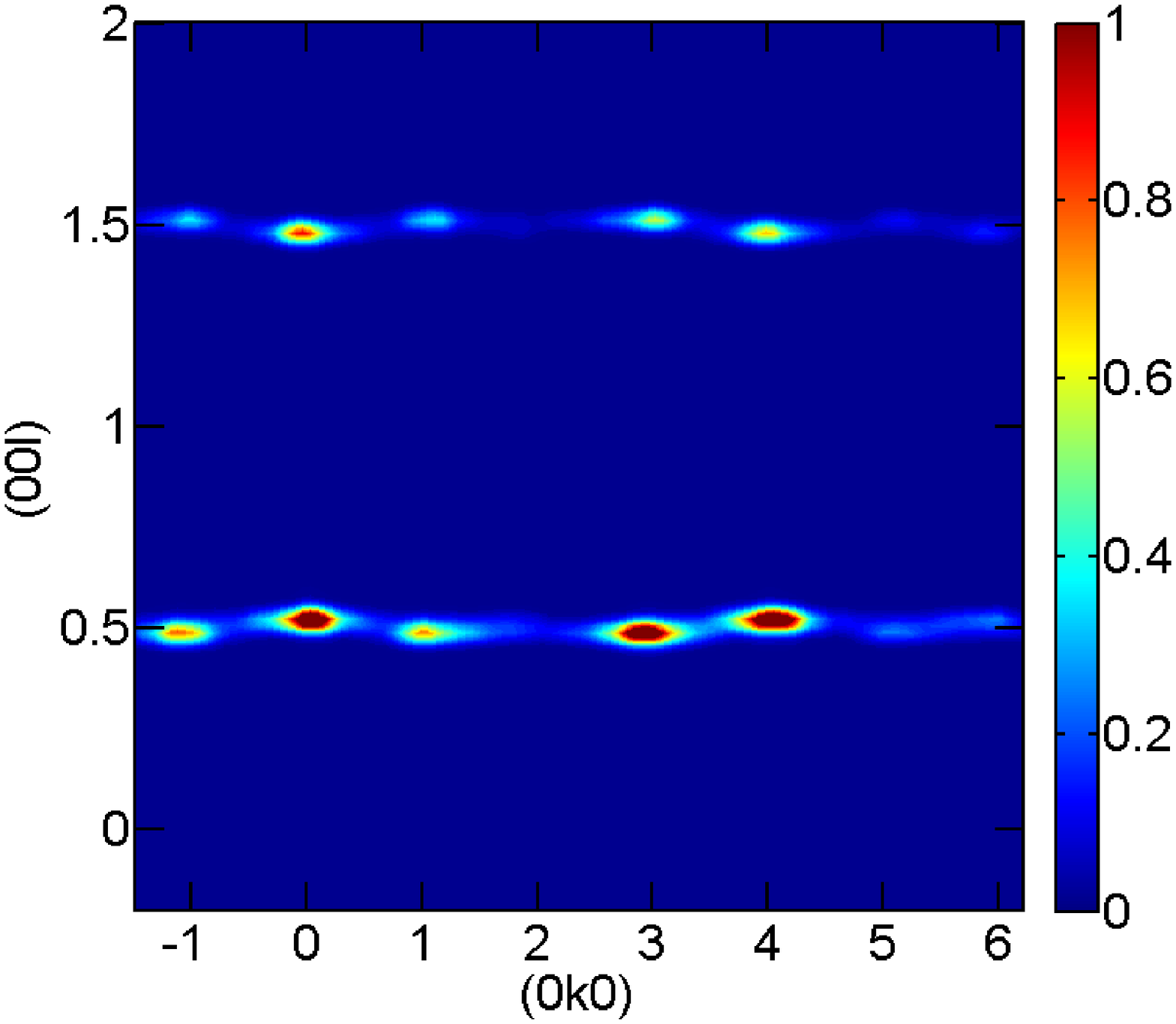}
\includegraphics[width={0.7\columnwidth}]{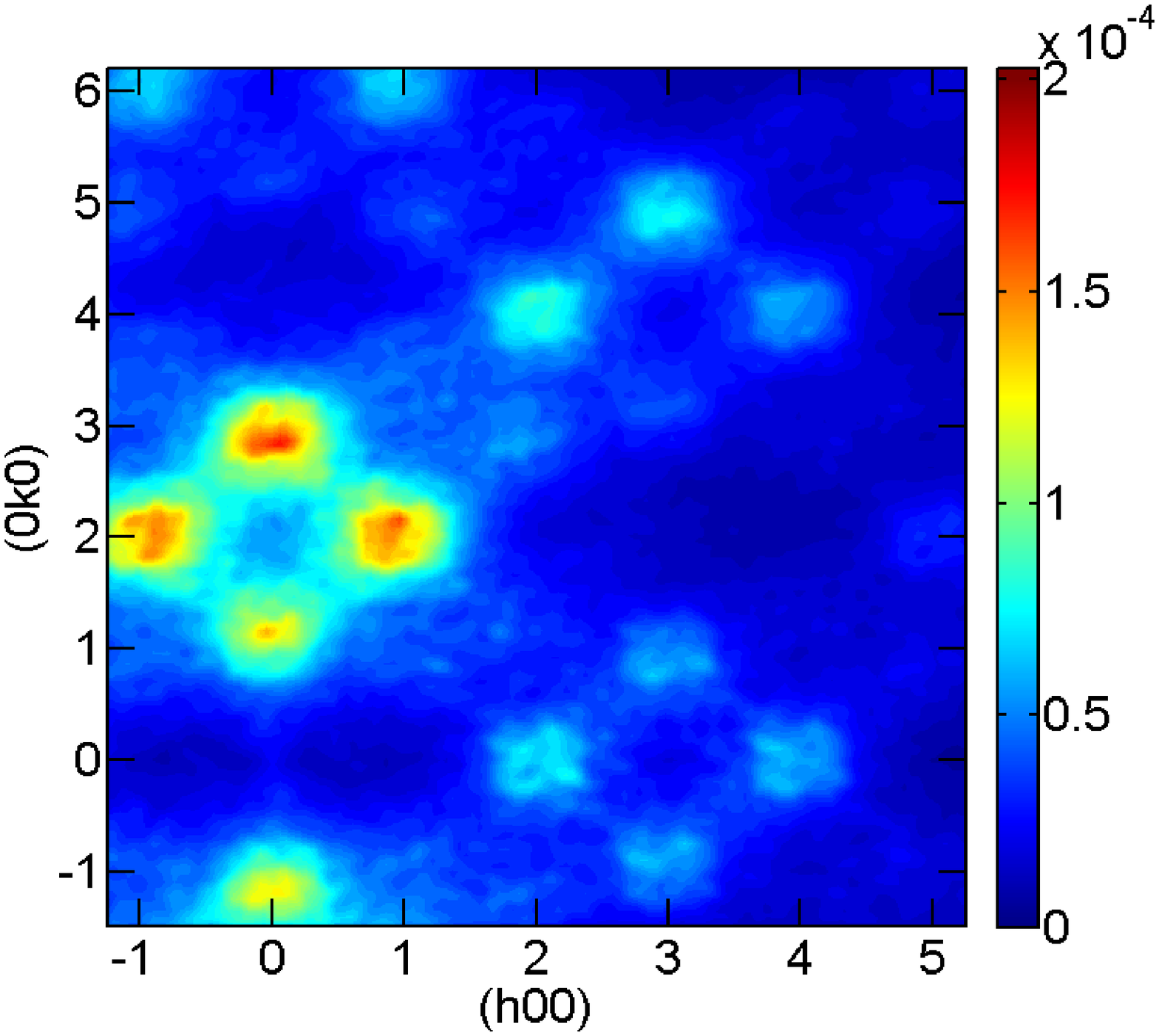}
\caption{(Colour online) Monte Carlo simulations of the reciprocal space maps of the magnetic scattering in the $(0kl)$ plane (top panel) and in the $(hk0)$ plane (bottom panel).
	     The intensity of magnetic scattering is calculated by averaging 10 different spin-configurations annealed from $T/J_1=400$ down to  $T/J_1=0.001$ (see text for detail).}
\label{Fig5_MC}
\end{figure}

Such parameterization produces a network of uncorrelated ladders where the spins are rotating in the \emph{a-c} plane in each ladder, and the associated scattering function displays noticeable modulation around the $l=\frac{1}{2}$ and $l=\frac{3}{2}$ positions in the $(0kl)$ scattering plane in broad agreement with the experimental data.
The correct modulation was obtained for a ratio $J_1/J_3\approx 5$ with both interactions made \afm\ in order to match the specific shape of the diffuse scattering rods in the $(0kl)$ plane.
This choice of the interaction signs makes sure that the scattering is more intense slightly above (below) the $l=\frac{1}{2}$ position for $k=0$ ($k=\pm 1$), but the situation is reversed around $l=\frac{3}{2}$. 

To progress further in our analysis, it is useful to note the equivalence of the well-studied linear chain model with nearest and next-nearest interactions\cite{chain_model} and the \ErII\ ladders described here if these were ``stretched'' along the $c$ axis.
The observed deviation of the most intense diffuse scattering in the $(0kl)$ plane from the $(0,0,1/2)$ position translates into a deviation of approximately 6 degrees from a perfect \afm\ arrangement of the nearest spins in the ladder.

Whilst this simple two-interaction model seems to capture most of the physics, some detail, such as the absence of scattering around the $k=2$ positions both for the $l=\frac{1}{2}$ and $l=\frac{3}{2}$ rods of scattering in the $(0kl)$ plane and the very peculiar pattern of intensity in the $(hk0)$ plane, are not properly accounted for.
In particular, the {\it lozenge} pattern of weak diffuse scattering, emphasised in the highlighted area in Fig.~\ref{Fig2_maps}a, is completely blurred in the simple $J_3$-$J_1$ model with the intensity not confined mostly to its corners.
This calls for the addition of another effective interaction between the ladders.

Keeping $J_1/J_3\approx 5$ constant, several such interactions were tried.
The best agreement between the simulated and experimentally observed patterns of diffuse $I(Q)$ has been obtained by an addition of the interaction denoted as $J^\prime$ in Fig.~\ref{Fig1_structure}.
Within the limitations of the minimalistic model used in the simulations, this interaction couples the magnetic ions belonging to different ladders and should be treated as an {\it effective} weak inter-ladder coupling rather than a specific exchange between ions in certain positions. 
The value of $J^\prime$ cannot be determined accurately, as its inclusion influences primarily the overall shape of the diffuse scattering rather than a particular feature whose position or intensity could be linked directly to the $J^\prime/J_1$ ratio.
Much more accurate experimental results as well as the simulations performed on much larger systems (to improve $Q$-resolution) will be required in order to establish $J^\prime$ reliably. 
The cuts through the simulated $I(Q)$ corresponding to the two measured scattering planes, $(hk0)$ and $(0kl)$, are presented in Fig.~\ref{Fig5_MC} for a model in which $J^\prime/J_1=0.01$.

\section{Discussion and Conclusions}
We return here to the intriguing issue of the dimensionality of the short-range correlations in \seo.
From an experimental point of view, in order to check whether the short-range correlations seen as undulated rods of intensity in Figs.~\ref{Fig2_maps}c and \ref{Fig2_maps}d are quasi-1D or 2D in nature, one has to make extended scans along the $a$ direction across the reciprocal space points such as $(0, 0, \frac{1}{2} + \delta)$, $(0, 0, \frac{3}{2} - \delta)$ or $(0, 4, \frac{1}{2} + \delta)$.
These types of scans are impossible on the neutron diffractometers used in this investigation without a complete realignment of the sample.
Given the neutron beam-time constraints the realignment was not performed, therefore the experimental information obtained so far is insufficient to draw a definite conclusion regarding the dimensionality of the correlations.

When carrying out the Monte Carlo simulations, it is straightforward to check whether the intensity observed as rods in the $(0kl)$ scattering plane in Fig.~\ref{Fig5_MC} protrudes significantly out of this plane.
Indeed, the results of our simulations suggest that this is the case and that the intensity forms undulated planes such as $(hk\frac{1}{2})$ and $(hk\frac{3}{2})$ rather than rods.
One has to bare in mind, however, that the simple model adopted in the simulations excludes the interactions between the two rare-earth sites. 
The results of the simulations call for further experimental studies, which should aim to probe the correlations along the $[100]$ direction in order to make a conclusive statement on this aspect of the magnetic ordering in \seo.

The onset of the fully ordered ${\bf k}=0$ structure at $T_N$ does not appear to affect the diffraction pattern of the diffuse phase.
In order to explain this lack of coupling between the two types of ordering, and the observation that a significant moment contributing to the ${\bf k}=0$ structure lies only on one of the two Er$^{3+}$ sites, we suggest that each of the sites contributes to only one of the two coexisting structures.
The different lengths of the inter-site exchange pathways (shown in light grey and dark grey in Fig.~\ref{Fig1_structure}) compared to those between symmetry equivalent positions of the same site may provide an explanation for the weak correlation between the two magnetic structures.

The low-dimensional character of the magnetic correlations could be a generic feature of several \slo\ compounds.
For example, in \sho\ Karunadasa {\it et al.}\cite{Karunadasa_2005} have reported the observation of broad magnetic peaks in powder neutron diffraction patterns at low temperature.
These peaks exhibit a characteristic Warren lineshape indicative of the presence of 2D ordering.\cite{Warren_1941}
It is not {\it a priori} clear, however, if on further cooling \sho\ will undergo a transition into a fully ordered 3D magnetic structure or (similarly to \seo) it will maintain separate 3D and low-dimensional magnetic components.
In another member of the same family of compounds, \sdo, no sharp magnetic Bragg peaks have been found down to 50~mK in the recent powder neutron diffraction experiments,\cite{Hayes_SrDyO} while the shapes of the observed broad diffuse-scattering peaks are not symmetric, which again implies 2D correlations.   

In conclusion, we have investigated the magnetic neutron diffraction patterns in single crystal \seo\ between 0.06 and 1.0~K, and have confirmed the presence of a 3D commensurate \afm\ state below 0.75~K.
We have also observed a short-range ordered magnetic phase that coexists with the commensurate phase, in which the size of the magnetic unit cell is nearly doubled along the $c$ axis and which does not exhibit critical behaviour.
The presence of two non-interacting magnetic phases in \seo\ is an example of the unusual physics arising from geometrically frustrated magnetism.

A classical Monte Carlo simulation shows that a simple model based on a ladder of triangles where the nearest neighbour interactions are approximately 5 times stronger than the next-nearest neighbour interactions mimics satisfactorily the observed diffuse scattering patterns. 
Even better agreement between the experimental results and simulations is achieved if the ladders are very weakly coupled through an effective further-neighbour interaction.

The authors are grateful for the support of the EPSRC, under grants EP/F02150X/1 and EP/E011802/1.
We also thank M.R.~Lees for a critical reading of the manuscript.


\begin{thebibliography}{100}
\bibitem{FWHMnote} The correlation length, $L$, is estimated using the expression: $\displaystyle L=\frac{2\pi}{\sqrt{W_{peak}^2-W_{ref}^2}}$, where $W_{peak}$ and $W_{ref}$ are the full widths at half maximum in Q-space for the Gaussian peak in question and a reference (resolution-limited) Bragg peak obtained on the same instrument respectively.
\bibitem{anisotropy} The actual value of the anisotropy term $D$ seems to have little influence on the results of the Monte Carlo simulations as long as it remains small compared to the exchange interactions $J_1$.
We have used the value $D/J_1=0.1$ throughout the simulations, but also checked that the value $D/J_1=0.01$ returns very similar results.
\end{thebibliography}
\end{document}